\documentclass[onecolumn,11pt]{IEEEtran}
\usepackage{cite}
\usepackage{amsmath,amssymb,amsfonts}
\usepackage{graphicx}
\usepackage{textcomp}
\usepackage{xcolor}
\def\BibTeX{{\rm B\kern-.05em{\sc i\kern-.025em b}\kern-.08em
    T\kern-.1667em\lower.7ex\hbox{E}\kern-.125emX}}
    \usepackage{tikz}
\usepackage[utf8]{inputenc}

\usepackage{amsmath} \usepackage{amsthm} \usepackage{amsfonts} \usepackage{amssymb} 
\usepackage{epstopdf}
\usepackage{graphicx}
\usepackage{bbm}

\usepackage{enumerate}

\newtheorem{theorem}{Theorem}

\newtheorem{lem}{Lemma}
\newtheorem{corollary}{Corollary}

\newtheorem{definition}{Definition}

\theoremstyle{definition}
\newtheorem{example}{Example}

\newtheorem*{prob*}{Problem}
\newtheorem{fact}{Fact}

\theoremstyle{remark}
\newtheorem{remark}{Remark}

\newcommand\numberthis{\addtocounter{equation}{1}\tag{\theequation}}

\global\long\def\RR{\mathbb{R}}

\global\long\def\NN{\mathbb{N}}

\global\long\def\ZZ{\mathbb{Z}}
\global\long\def\EE{\mathbb{E}}

\begin{document}

\title{On The Reliability Function of Discrete Memoryless Multiple-Access Channel with Feedback}
\author{
 \IEEEauthorblockN{Mohsen Heidari, Achilleas Anastasopoulos, and S. Sandeep Pradhan}
%
%
%
\thanks{The authors are with the Department of Electrical Engineering and Computer Science, University of Michigan, Ann Arbor, MI, 48105 USA e-mail: {mohsenhd,
anastas, pradhanv}@umich.edu}
}
\IEEEoverridecommandlockouts

\maketitle
\begin{abstract}
We derive a lower and upper bound on the reliability function of discrete memoryless multiple-access channel (MAC) with noiseless feedback and variable-length codes (VLCs). For the upper-bound, we use proof techniques of Burnashev for the point-to-point case. Also, we adopt the techniques used to prove the converse for the feedback-capacity of MAC. For the lower-bound on the error exponent, we present a coding scheme consisting of a data and a confirmation stage.
In the data stage, any arbitrary feedback capacity-achieving code is used. In the confirmation stage, each transmitter sends one bit of information to the receiver using a pair of codebooks of size two, one for each transmitter. The codewords at this stage are selected randomly according to an appropriately optimized joint probability distribution.  The bounds increase linearly with respect to a specific Euclidean distance measure defined between the transmission rate pair and the capacity boundary. The lower and upper bounds match for a class of MACs.     
\end{abstract}

\section{Introduction}
Noiseless feedback does not increase the capacity for communications over discrete memoryless channels (DMC) \cite{Shannon_zero}. Furthermore, Dobrushin \cite{Dobrushin} and later Haroutunian \cite{Haroutunian} showed that feedback does not improve the error exponent of symmetric channels  when fixed-length codes are used. 
Nevertheless,  feedback can be very useful in the context of variable-length codes. 

In a remarkable work, Burnashev \cite{Burnashev} demonstrated that the error exponent improves for DMCs with feedback and variable-length codes.  The error exponent has a simple form
\begin{align}\label{eq: E(R) error exponent }
E(R)=(1-\frac{R}{C}) C_1,
\end{align}
where $R$ is the (average) rate  of transmission, $C$ is the capacity of the channel, and $C_1$ is  the maximal relative entropy between conditional output distributions. Berlin et al \cite{Berlin} have provided a simpler derivation of the Burnashev bound that emphasizes the link between the constant $C_1$ and the binary hypothesis testing problem. Yamamoto and Itoh \cite{Yamamoto} introduced a coding scheme that its error exponent achieves $E(R)$ in \eqref{eq: E(R) error exponent }. Their scheme consists of two distinct transmission phases that we called the data and the confirmation phase, respectively. In the data stage the message is encoded using a capacity achieving fixed blocklength code.  During the confirmation phase, the transmitter
sends one bit of information to the receiver. The decoder performs a binary hypothesis test to decide if $0$ or $1$ is transmitted.

In the context of communications over multi-user channels, the benefits of feedback are more prominent. For instance, Gaarder and Wolf \cite{Gaarder-Wolf} showed that feedback can expand the capacity region of discrete memoryless multiple-access channels (MAC). Willems \cite{Willems-FB} derived the feedback-capacity region for a class of MACs. Characterizing the capacity region and the error exponent for general MACs remains an open problem. Using \textit{directed information} measures, Kramer \cite{Kramer-thesis} was able to characterize the feedback-capacity region of two-user MAC with feedback. However, the characterization is in the form of infinite letter directed information measures which is not computable in general. The error exponent for discrete memoryless MAC without feedback is studied in \cite{nazari-MAC-FB-UB, Gallager}.

In this paper, we study the error exponent of discrete memoryless MAC with noiseless feedback. In particular, we derive an upper-bound and a lower-bound. For that, let $(||\underline{R}||, \theta_R)$ denote the polar coordinate of $(R_1,R_2)$ in $\RR^2$. In this setting, the upper-bound is
\begin{align}\label{eq: E_u polar}
E_u(R_1,R_2)=(1-\frac{||\underline{R}||}{C(\theta_R)}) D_u
\end{align}
where $C(\theta_R)$ is the point of the capacity frontier at the angle determined by $\underline{R}$. The lower-bound is the same as $E_u$ but with different constant $D_l$. The constants $D_l$ and $D_u$ are determined by the relative entropy between the conditional output distributions. We show that for a class of MACs the two bounds coincide. 

The paper is organized as follows: In Section \ref{sec: problem formulation}, basic definitions and the problem formulation are provided. In Section \ref{sec: lower bound}, we derive a lower-bound for the reliability function. In Section \ref{sec: upper bound on exp}, we characterize an upper-bound for the reliability function. In Section \ref{sec: the shape of the bounds}, we compare the lower and upper-bound and explore examples for the tightness of the bounds. Finally, Section \ref{sec: conclution} concludes the paper. 

\section{Problem Formulation and Definitions}\label{sec: problem formulation}

Consider a discrete memoryless MAC with input alphabets $\mathcal{X}_1,\mathcal{X}_2$, and output alphabet $\mathcal{Y}$. The channel conditional probability distribution is denoted by $Q(y|x_1, x_2)$ for all $(y, x_1, x_2)\in \mathcal{Y}\times \mathcal{X}_1\times \mathcal{X}_2$. Such setup is denoted by $(\mathcal{X}_1,\mathcal{X}_2, \mathcal{Y}, Q)$.  Let $y^t$ and $x_{i}^t$, $i=1,2,$ be the channel output and the inputs sequences after $t$ uses of the channel, respectively. Then, the following condition is satisfied:
\begin{align}\label{eq: chann probabilities}
P(y_t|y^{t-1}, x_1^{t-1},x_2^{t-1})=Q(y_t|x_{1t}, x_{2t}).
\end{align}
We assume that the output of the channel as a feedback is available at the encoders with one unit of delay. 
\begin{definition}
An $(M_1, M_2, N)$- variable-length code (VLC) for a MAC $(\mathcal{X}_1,\mathcal{X}_2, \mathcal{Y}, Q)$ with feedback is defined  by 
\begin{itemize}
\item A pair of messages $W_1, W_2$ selected randomly with uniform distribution from  $ \{1,2, \dots, M_i\}, i=1,2$.


\item Two sequences of encoding functions 
\begin{align*}
e_{i,t}: \{1,2, \dots, M_i\} \times \mathcal{Y}^{t-1} \rightarrow \mathcal{X}_i, \quad t\in \NN, ~ i=1,2,
\end{align*}
one for each transmitter. 
\item A sequence of decoding functions 
\begin{align*}
d_t: \mathcal{Y}^t \rightarrow \{1,2, ... , M_1\} \times \{1,2, ... , M_2\}, \quad t\in \NN.
\end{align*}
\item  A stopping time $T$ with respect to (w.r.t) the filtration $\mathcal{F}_t$ defined as the $\sigma$-algebra of $Y^t$ for $t\in \NN$. Furthermore, it is assumed that $T$ satisfies $\EE[T] \leq N$.
\end{itemize}

\end{definition}
For each $i=1,2$, given a message $W_i$, the $t$th output of Transmitter $i$ is denoted by $X_{i,t}=e_{i,t}(W_i, Y^{t-1})$.

 Let $(\hat{W}_{1,t}, \hat{W}_{2,t})=d_t(Y^t)$. Then, the decoded messages at the decoder are denoted by $\hat{W}_1=\hat{W}_{1,T}$, and $\hat{W}_2=\hat{W}_{2,T}$. In what follows, for any $(M_1, M_2, N)$ VLC, we define average rate-pair, error probability, and error exponent. 
Average rates for an $(M_1, M_2, N)$ VLC are defined as  
 $$R_i \triangleq \frac{\log_2 M_i}{\EE [T]},\quad  i=1,2.$$ 

The probability of error is defined as $$P_e =P\left((\hat{W}_1,\hat{W_2} )\neq (W_1,W_2) \right).$$  

The error exponent of a VLC with probability of error $P_e$ and stopping time $T$ is defined as $E\triangleq-\frac{\log_2 P_e}{\EE[T]}$.
\begin{definition}
A reliability function $E(R_1,R_2)$ is said to be achievable for a given MAC, if for any $R_1, R_2>0$ and $\epsilon>0$ there exists an $(M_1,M_2, N)$-VLC such that
 \begin{align*}
 -\frac{\log_2 P_e}{N} \geq E(R_1, R_2)-\epsilon, ~ \text{and}~\frac{\log_2 M_i}{N} \geq R_i -\epsilon,
 \end{align*}
 where $i=1,2$, and $P_e$ is the error probability of the VLC.
\end{definition}

\begin{definition} 
The reliability function of a MAC with feedback is defined as the supremum of all achievable reliability functions $E(R_1,R_2)$. 
\end{definition}

\subsection{The Feedback-Capacity Region of MAC}\label{subsec: feedback capacity of MAC}
We summarize Kramer's results presented in  \cite{Kramer-thesis} for the feedback capacity of MAC. 
 We use \textit{directed information} and \textit{conditional directed information} as defined in  \cite{Kramer-thesis}. The normalized directed information from a sequence $\mathbf{X}^n$ to a sequence $\mathbf{Y}^n$ when causally conditioned on $\mathbf{Z}^n$ is denoted by 
\begin{align}\label{eq: normalized direct info}
I_n(X\rightarrow Y|| Z)=\frac{1}{n}I(\mathbf{X}^n \rightarrow \mathbf{Y}^n|| \mathbf{Z}^n).
\end{align}
%
%
The feedback-capacity region of a discrete memoryless MAC with feedback $(\mathcal{X}_1,\mathcal{X}_2, \mathcal{Y}, Q)$ is denoted by $\mathcal{C}$, and is the closure of the set of all rate-pairs $(R_1, R_2)$ such that  
\begin{align*}
R_1 &\leq I_L(X_1 \rightarrow Y|| X_2)\\
R_2 &\leq I_L(X_2 \rightarrow Y|| X_1)\\
R_1+R_2 &\leq I_L(X_1 X_2 \rightarrow Y),
\end{align*}
where $L$ is a positive integer, and $P_{X_1^LX_2^LY^L}$ factors as 
\begin{align}\label{eq: factorization of P}
\prod_{l=1}^L P_{1,l}(x_{1l}| x_1^{l-1} y^{l-1})P_{2,l}(x_{2l}| x_2^{l-1} y^{l-1})Q(y_l|x_{1, l}x_{2,l}).
\end{align}

\begin{definition}\label{def: C lambda}
Let $\lambda_1, \lambda_2, \lambda_3 \geq 0$, and $\lambda_1+\lambda_2+\lambda_3=1$. Define 
\begin{align*}
C_{\underline{\lambda}}=\sup_{L \in \NN} \sup_{P_{X_1^LX_2^LY^L}} & \lambda_1I_L(X_1\rightarrow Y|X_2)+\lambda_2I_L(X_2\rightarrow Y|X_1)\\ &+\lambda_3I_L(X_1 X_2\rightarrow Y),
\end{align*}
where $P_{X_1^LX_2^LY^L}$ factors as in  \eqref{eq: factorization of P}.
\end{definition}

\begin{fact}\label{Fact: MAC-FB capacity hyperplane}
The feedback-capacity of a discrete memoryless MAC with feedback is the same as the closure of the set of rate-pairs $(R_1, R_2)$ such that the inequality
\begin{align*}
\lambda_1 R_1+\lambda_2 R_2 +\lambda_3 (R_1+R_2) \leq C_{\underline{\lambda}}
\end{align*}
holds for all  $\lambda_1, \lambda_2, \lambda_3 \geq 0$, with $\lambda_1+\lambda_2+\lambda_3=1$.
\end{fact}

\subsection{Notational Conventions}
For more convenience, we denote a rate-pair $(R_1,R_2)$ by $(R_1,R_2,R_3)$, where $R_3=R_1+R_2$. For a $(\mathcal{X}_1, \mathcal{X}_2, \mathcal{Y},Q)$ MAC we use the following notational convenience 
\begin{align}\label{eq: I_L^1}
I_L^1&\triangleq I_L(X_1\rightarrow Y||X_2),\\\label{eq: I_L^2}
I_L^2&\triangleq I_L(X_2\rightarrow Y||X_1),\\\label{eq: I_L^3}
I_L^3&\triangleq I_L(X_1 X_2\rightarrow Y).
\end{align}
The Kullback–Leibler divergence for the MAC with transition probability matrix $Q$ is defined as
\begin{align*}
 D_Q(x_1,x_2||z_1,z_2)= \sum_{y\in \mathcal{Y}} Q(y|x_1,x_2) \log_2 \frac{Q(y|x_1,x_2)}{Q(y|z_1,z_2)},
\end{align*}
where $(x_1,x_2), (z_1,z_2) \in \mathcal{X}_1 \times \mathcal{X}_2$. For notational convenience we denote 
\begin{align*}
D_1(x_1,x_2|| z_1,z_2)&=  D_Q(x_1,x_2||z_1, x_2)\\
D_2(x_1,x_2|| z_1,z_2)&=  D_Q(x_1,x_2||x_1, z_2)\\
D_3(x_1,x_2|| z_1,z_2)&=  D_Q(x_1,x_2||z_1, z_2).
\end{align*}

\section{A Lower-Bound for the Reliability Function}\label{sec: lower bound}
%


We build upon Yamamoto-Itoh transmission scheme for point-to-point (ptp) channel coding with feedback \cite{Yamamoto}. The scheme sends the messages $W_1, W_2$ through blocks of length $n$. The transmission process is performed in two stages: 1) The ``data transmission" stage taking up to $n(1-\gamma)$ channel uses, 2) The ``confirmation" stage taking up to $n\gamma$ channel uses, where $\gamma$ is a design parameter taking values from  $[0,1]$.

\paragraph*{\textbf{Stage 1}}
For the first stage, we use any coding scheme that achieves the feedback-capacity of the MAC. The length of this coding scheme is at most $n(1-\gamma)$.  Let $\hat{W}_1, \hat{W}_2$ denote the decoder's estimation of the messages at the end of the first stage. Define the following random variables:
\begin{align*}
H_i=1\{\hat{W}_i \neq W_i\}, \quad i=1,2.
\end{align*}
Because of the feedback, $\hat{W}_1$ and $\hat{W}_2$ are known at each transmitter. 
Therefore, at the end of the first stage, transmitter $i$ has access to $W_i, \hat{W}_1, \hat{W}_2$, and $H_i$, where $i=1,2$.  

\paragraph*{\textbf{Stage 2}}
The objective of the second stage is to inform the receiver whether the hypothesis $\Theta_0: (\hat{W}_1, \hat{W}_2)=(W_1,W_2)$ or  $\Theta_1: (\hat{W}_1, \hat{W}_2)\neq (W_1,W_2)$ is correct. 
For that, each transmitter employs a code of size two and length $\gamma n$. The codewords of such codebooks are denoted by two pairs of sequences $(\underline{x_1}(0),\underline{x_2}(0))$ and $(\underline{x_1}(1), \underline{x_2}(1))$ each with elements belonging to $\mathcal{X}_1 \times \mathcal{X}_2$. Fix a joint-type $\mathsf{P}_n$ defined over the set $\mathcal{X}_1 \times \mathcal{X}_2\times \mathcal{X}_1 \times \mathcal{X}_2$ and for sequences of length $\gamma n$. The sequences  $(\underline{x_1}(0),\underline{x_2}(0), \underline{x_1}(1), \underline{x_2}(1))$ are selected randomly among all the sequences with joint-type $\mathsf{P}_n$.  During this stage and given $H_1$, Transmitter $1$ sends $\underline{x_1}(H_1)$. Similarly, Transmitter 2 sends $\underline{x_2}(H_2)$.

\paragraph*{\textbf{Decoding}}
Upon receiving the channel output, the receiver estimates $H_1, H_2$. Denote this estimation by $\hat{H}_1, \hat{H}_2$. 
If $(\hat{H}_1, \hat{H}_2) = (0,0)$, then the hypothesis $\hat{\Theta}=\Theta_0$ is declared. Otherwise, $\hat{\Theta}=\Theta_1$ is declared. Because of the feedback, $\hat{\Theta}$ is also available at each encoders. If $\hat{\Theta}=\Theta_0$, then transmission stops and a new data packet is transmitted at the next block. Otherwise, the message is transmitted again at the next block. The process continues until $\hat{\Theta}=\Theta_0$ occurs. 

The confirmation stage in the proposed scheme can be viewed as a decentralized binary hypothesis problem in which a binary hypothesis $\{\Theta_0, \Theta_1\}$ is observed partially by two distributed agents and the objective is to convey the true hypothesis to a central receiver. This problem is qualitatively different from the sequential binary hypothesis testing problem as identified in \cite{Berlin} for ptp channel. Note also that in the confirmation stage we use a different coding strategy than the one used in Yamamoto-Itoh scheme \cite{Yamamoto}. Here, all four codewords have a joint-type  $\mathsf{P}_n$. It can be shown that repetition codes, and more generally, constant composition codes are strictly suboptimal in this problem.  

\begin{theorem}\label{thm: Error Exp lowerbound}
The following is a lower-bound for the reliability function  of any discrete memoryless MAC:
\begin{align}\label{eq: E_l}
E_l(R_1,R_2) = \min_{\substack{\lambda_1, \lambda_2, \lambda_3 \geq 0\\ \lambda_1+\lambda_2+\lambda_3=1}} D_l (1-\frac{\sum_i \lambda_i R_i}{C_{\underline{\lambda}}}) ,
\end{align}
where, 	
\begin{align}\label{eq: D_l}
D_l\triangleq \sup_{{P_{X_1X_2Z_1Z_2}}}\min_{i=1,2,3} \EE\left[D_i(X_1,X_2||Z_1,Z_2)\right],
\end{align}
and the supremum is taken over all probability distributions $P_{X_1X_2Z_1Z_2}$ defined over $\mathcal{X}_1 \times
\mathcal{X}_2 \times
\mathcal{X}_1 \times
\mathcal{X}_2
$. 
\end{theorem}
\begin{IEEEproof}
The proof is given in Appendix \ref{appx: proof of thm error Exp lowebound}.
\end{IEEEproof}



\section{An Upper-bound for the Reliability Function}\label{sec: upper bound on exp}

In this part of the paper, we establish an upper-bound for the reliability function of any discrete memoryless MAC.
Define
\begin{align}\label{eq: definiton of D_i}
D_i&\triangleq \max_{\substack{x_1, z_1\in \mathcal{X}_1,\\ x_2,z_2 \in \mathcal{X}_2}} D_i(x_1,x_2||z_1, z_2), \quad i=1,2,3.
\end{align}

\begin{theorem}[Upper-bound]\label{thm: Error Exp Upper bound}
For any $(N, M_1,M_2)$ VLC with probability of error $P_e$, and any $\epsilon>0$, there exists a function $\delta$ such that the following is an upper-bound for the reliability function of the VLC
\begin{align}\nonumber
E(R_1,R_2)\leq &\min_{\substack{\lambda_1, \lambda_2, \lambda_3 \geq 0\\ \lambda_1+\lambda_2+\lambda_3=1}}\min_{j\in \{1,2,3\}} D_j \left(1- \frac{\lambda_j R_j}{C_{\lambda}}\right)\\\label{eq: E_u} &+\delta(P_e, M_1M_2, \epsilon),
\end{align}
where $(R_1, R_2)$ is the rate pair of the VLC and $\delta$ satisfies $$\lim_{\epsilon\rightarrow 0} \lim_{P_e\rightarrow 0}\lim_{M_1M_2\rightarrow \infty} \delta(P_e,M_1M_2, \epsilon)=0.$$
\end{theorem}

\begin{corollary}\label{cor: Exp upper bound}
From Theorem \ref{thm: Error Exp Upper bound}, the following is an upper-bound for the error exponent of a MAC: 
\begin{align}\nonumber
E_u(R_1,R_2)=& \min_{\substack{\lambda_1, \lambda_2, \lambda_3 \geq 0\\ \lambda_1+\lambda_2+\lambda_3=1}}D_{u} \left(1-\frac{\sum_{i=1}^3  \lambda_i R_i}{C_{\lambda}}\right)+\delta,
\end{align}
where $D_u=\max\{D_1, D_2, D_3\}$, and $\delta$ is as in Theorem \ref{thm: Error Exp Upper bound}.
\end{corollary}
\begin{IEEEproof}
The proof is given in Appendix \ref{app: proof corrolary Exp upper bound}. 
\end{IEEEproof}

\subsection{Proof of the Upper-Bound}\label{subsec: proof of theorem}
Consider any $(N, M_1,M_2)$ VLC with probability of error $P_e$, and stopping time $T$. Suppose the message at Encoder 2, $W_2$, is made available to all terminals. For the new setup, as $W_2$ is available at the Decoder, the average probability of error is $P_e^1 \triangleq P\{\hat{W}_1\neq W_1\}$. Note that $P_e\geq P_e^1$. We refer to such setup as $W_2$-assisted MAC. For a maximum a \textit{posteriori} decoder, after $n$ uses of the channel and assuming the realization $Y^n=y^n$ and $W_2=w_2$, define
\begin{align*}
T_1^{\delta} \triangleq \inf \big\{n: \max_{1\leq i \leq M_1} P(W_1=i| y^n, w_2)\geq 1-\delta \big \},
\end{align*}
where $\delta>0$ is a fixed real number. Also, let $\tau_1 \triangleq \min  \{ T, T_1^{\delta}\}$. Note that $\tau_1$ is a stopping time w.r.t the filtration $\{\mathcal{F}_{W_2}\times \mathcal{F}_t\}_{t>0}$. The following lemma provides a lower-bound on the probability of error for such setup. 

\begin{lem}\label{lem: hypothesis gini-aided MAC}
The probability of error, $P_e$, for a hypothesis testing over a $W_2$-assisted MAC and variable length codes satisfies the following inequality 
\begin{align*}
P_e\geq \frac{\min \{P(H), P(H^c)\}}{4}e^{-D_1\EE[T]},
\end{align*}  
where $\{H, H^c\}$ are the two hypothesizes and $T$ is the stopping time of the variable length code. 
\end{lem}

\begin{lem}\label{lem: zeta}
For a given MAC with finite $D_3$ the following holds
\begin{align*}
\zeta p(w_1, w_2|y^{n-1})\leq  p(w_1, w_2|y^{n})\leq \frac{ p(w_1, w_2|y^{n-1})}{\zeta},
\end{align*}
where $\zeta\triangleq \min_{x_1,x_2,y} Q(y|x_1,x_2)$.
\end{lem}
The above lemmas are extensions of Lemma 1 and Proposition 2 in \cite{Berlin} for MAC. The proofs follow from similar arguments and are omitted.

\begin{lem}\label{lem: P_e lowerbound W_2 assisted MAC}
Given a MAC with $D_3<\infty$, and for any $(N, M_1,M_2)$ VLC with probability of error $P_e$ the following holds
\begin{align}\label{eq: P_e^1}
P_e \geq \frac{\zeta \delta}{4} e^{-D_1 \EE[T- \tau_1]},
\end{align}
where $\zeta\triangleq \min_{x_1,x_2,y} Q(y|x_1,x_2)$.
\end{lem} 

\begin{IEEEproof}
Suppose the VLC is used for a $W_2$-assisted MAC. As discussed before, $P_e\geq P_e^1$. 
%
We modify the encoding and the decoding functions of the VLC used for the MAC. Let $\mathcal{H}_1 \subseteq \mathcal{M}_1$ be a subset of the message set $\mathcal{M}_1$.  The subset $\mathcal{H}_1$ is to be determined at time $\tau_1$. The new decoding function, at time $T$, decides whether the message belongs to $\mathcal{H}_1$. The new encoding functions are the same as the original one until the time $\tau_1$. Then, after $\tau_1$, the transmitters perform a VLC to resolve the binary hypothesis  $\{W_1\in \mathcal{H}_1\}$ and $\{W_1\notin \mathcal{H}_1\}$. This hypothesis problem is performed from $\tau_1$ to $T$. With these modifications, the error probability of this binary hypothesis problem is a lower-bound on $P_e$.  In what follows, we present a construction for $\mathcal{H}_1$. Then, we apply Lemma \ref{lem: hypothesis gini-aided MAC} to complete the proof. 

Let $P_e^1(y^n, w_2)\triangleq 1-\max_{1\leq i \leq M_1} P(W_1=i| y^n, w_2).$ 
 The quantity $P_e^1(y^{\tau_1}, w_2)$ can be calculated at all terminals. 
  By definition, at time $\tau_1-1$, the inequality $P(W_1=i| Y^{\tau_1-1}, W_2)<1-\delta$ holds almost surely for all $i\in [1:M_1]$.  This implies that $P_e^1(Y^{\tau_1-1}, W_2)>\delta$. Hence, by Lemma \ref{lem: zeta}  at time $\tau_1$ the inequality $P_e^1(Y^{\tau_1},W_2)\geq \zeta \delta$ holds almost surely.  We consider two cases $P_e^1(y^{\tau_1}, w_2)\leq \delta$ and $P_e^1(y^{\tau_1}, w_2)>\delta$, where $\delta$ is the constant used in the definition of $T_1^{\delta}$.  For the first case,  $\mathcal{H}_1$ is the set consisting of the message with the highest a \textit{posteriori} probability. Since $P_e^1(y^{\tau_1}, w_2)\leq \delta$, then $P(\mathcal{H}_1)\geq 1-\delta$. In addition, as $P_e^1(y^{\tau_1}, w_2)\geq \zeta \delta$, then $P(\mathcal{H}^c_1)>\zeta \delta$. For the second case, set $\mathcal{H}_1$ to be a set of messages such that $P(\mathcal{H}_1)>\delta/2$ and $P(\mathcal{H}_1)<1-\delta$. Such set exists, since $P(W_1=i| Y^{\tau-1}, W_2)<1-\delta$ holds for all messages $i\in [1:M_1]$.

Note that by the above construction, for each case, $P(\mathcal{H}_1) \in [\zeta \delta, 1-\zeta\delta] $. Thus, from Lemma \ref{lem: hypothesis gini-aided MAC} and the argument above, the inequality 
\begin{align*}
P\{\hat{W}_1\neq W_1|Y^{\tau},W_2\} \geq \frac{\zeta \delta}{4} e^{-D_1 \EE[T- \tau|Y^{\tau}, W_2]}
\end{align*}
holds almost surely. Next, we take the expectation of the above expression. The lemma follows by the convexity of $e^{-x}$ and Jensen's inequality. 

\end{IEEEproof}
Next, we apply the same argument for the case where $W_1$ is available at all the terminals. For that define
\begin{align*}
T_2^{\delta}&\triangleq \inf \big\{n: \max_{1\leq j \leq M_2} P(W_2=j| y^n, w_1)\geq 1-\delta \big \},
\end{align*}
and let $\tau_2 \triangleq \min  \{ T, T_2^{\delta}\}$. By symmetry, Lemma \ref{lem: P_e lowerbound W_2 assisted MAC} holds for this case and we obtain
\begin{align}\label{eq: P_e^2}
P_e \geq \frac{\zeta \delta}{4} e^{-D_2 \EE[T- \tau_2]}.
\end{align}
Next, define the following stopping times:
\begin{align*}
T_3^{\delta}&\triangleq \inf \big\{n: \max_{i,j} P(W_1=i, W_2=j| y^n)\geq 1-\delta \big \}.
%
\end{align*}
Also, let $\tau_3=\min\{T, T_3^{\delta}\}$. using a similar argument as in the above, we can show that 
\begin{align}\label{eq: P_e^3}
P_e \geq \frac{\zeta \delta}{4} e^{-D_3 \EE[T- \tau_3]}.
\end{align}
For that, after time $\tau_3$, we formulate a binary hypothesis problem in which the transmitters determine whether $(W_1,W_2)\in \mathcal{H}_3$ or not. Here, $\mathcal{H}_3$ is a subset which is constructed using a similar method as for $\mathcal{H}_1$ in the proof of Lemma \ref{lem: P_e lowerbound W_2 assisted MAC}.  We further allow the transmitters to communicate with each other after $\tau_3$. 
The maximum of the right-hand sides of \eqref{eq: P_e^1}, \eqref{eq: P_e^2} and \eqref{eq: P_e^3} gives a lower-bound on $P_e$. The lower-bound depends on the expectation of the stopping times $\tau_i, i=1,2,3$. In what follows, we provide a lower-bound on $\EE[\tau_i]$. 
 Define the following random processes.
 \begin{equation*}
 \begin{aligned}[c]
H^1_t&\triangleq H(W_1|~\mathcal{F}_{W_2}\times \mathcal{F}_t),\\ 
H^2_t &\triangleq H(W_2|~\mathcal{F}_{W_1}\times \mathcal{F}_t),\\
H^3_t&\triangleq H(W_1,W_2|~ \mathcal{F}_t),
\end{aligned}
 \end{equation*}
\begin{lem}\label{lem: linear drift MAC FB}
Given a $(M_1, M_2, N)$-VLC, for any $\epsilon>0$ there exist $L$ and a probability distribution $P_{X_1^LX_2^LY^L}$ that factors as in \eqref{eq: factorization of P} such that 
the following inequalities hold almost surely for $1 \leq t \leq N$
\begin{align*}
\EE[H^1_{t+1}-H^1_{t}|\mathcal{F}_{W_2}\times \mathcal{F}_t] & \geq - (I^1_L+\epsilon), \\
\EE[H^2_{t+1}-H^2_{t}|\mathcal{F}_{W_1}\times \mathcal{F}_t] & \geq - (I^2_L+\epsilon),\\
\EE[H^3_{t+1}-H^3_{t}|\mathcal{F}_t] & \geq - (I^3_L+\epsilon).
\end{align*}
where $i=1,2,3$, and $I_L^i$ is defined as in \eqref{eq: I_L^1}-\eqref{eq: I_L^3}.
\end{lem}
\begin{IEEEproof}
The proof is provided in Appendix \ref{appx: proof of lem linear drift}.
\end{IEEEproof}
We need the following lemma to proceed. The lemma is a result of Lemma 4 in \cite{Burnashev}, and we omit its proof.
\begin{lem}\label{lem: H_t+1 - H_tis bounded}
For any $t\geq 1$ and $i=1,2,3$, the following inequality holds almost surely w.r.t $\mathcal{F}_{W_1}\times \mathcal{F}_{W_2} \times \mathcal{F}_t$
\begin{align*}
\log H_t^i-\log H_{t+1}^i \leq \max_{\substack{j,l\in [1:M_1]\\ k,m\in [1:M_2]}}\max_{y\in \mathcal{Y}} \frac{\hat{Q}_{j,k}(y)}{\hat{Q}_{l,m}(y)}.
\end{align*}
\end{lem}

From Lemma \ref{lem: linear drift MAC FB} and the fact that $H_t^i\leq \log_2 M_i < \infty$, the processes $\{H_t^i+ (I_L^1+\epsilon)t\}_{t>0}$ are submartingales for $i=1,2,3$. In addition, from Lemma \ref{lem: H_t+1 - H_tis bounded} and the inequalities $\EE[\tau_i]\leq \EE[T]\leq N <\infty$, we can apply Doob's Optional Stopping Theorem for each submartingale $\{H_t^i+ (I_L^1+\epsilon)t\}_{t>0}$. Then, we get:
\begin{align}\label{eq: bound on log M_i for MAC Exp}
\log M_i &\leq \EE[H^i_{\tau_i}]+\EE[\tau_i] (I_L^i+\epsilon)
\end{align}
where $M_3=M_1 M_2$.

\begin{lem}
The following inequality holds for each $i=1,2,3$
\begin{align*}
\EE[H^i_{\tau_i}]\leq h_b(\delta)+(\delta+\frac{P_e}{\delta})\log_2 M_i.
\end{align*}
\end{lem}
\begin{IEEEproof}
We prove the lemma for the case $i=1$. The proof for $i=2,3$ follows from a similar argument. For $i=1$, we obtain 
\begin{align}\nonumber
\EE[H_{\tau_1}^1]&=P\{P_e(Y^{\tau_1},W_2)>\delta\}\EE[H_{\tau_1}^i|P_e(Y^{\tau_1},W_2)>\delta]+P\{P_e(Y^{\tau_1},W_2)\leq \delta\}\EE[H_{\tau_1}^1|P_e(Y^{\tau_1},W_2)\leq \delta]\\\label{eq: bound on EE H_tau_i}
&\leq P\{P_e(Y^{\tau_1},W_2)>\delta\}\log_2 M_1 + P\{P_e(Y^{\tau_1},W_2)\leq \delta\}\EE[H_{\tau_1}^i|P_e(Y^{\tau_1},W_2)\leq \delta].
\end{align}

Note that the event $\{P_e(Y^{\tau_1},W_2)>\delta\}$ implies that $\tau_1=T$, and $P_e(y^{\tau_1},W_2)>\delta$ for all $0\leq n \leq T$. Hence, this event is included in the event $\{P_e(Y^{T},W_2)>\delta\}$. Thus, applying Markov inequality gives 
\begin{align*}
P\{P_e(Y^{\tau_1},W_2)>\delta\}\leq P\{P_e(Y^{T},W_2)>\delta\} \leq \frac{P_e}{\delta}.
\end{align*}
As a result of the above argument, the right-hand side of \eqref{eq: bound on EE H_tau_i} does not exceed the following 
 \begin{align*}
 \frac{P_e}{\delta} \log_2 M_1 + \EE[H_{\tau_1}^1|P_e(Y^{\tau_1},W_2)\leq \delta].
 \end{align*}
 From Fano's inequality we obtain 
\begin{align*}
\EE[H_{\tau_1}^1 | P_e(Y^{\tau_1},W_2)\leq \delta]\leq h_b(\delta)+\delta \log_2 M_1.
\end{align*}
The proof is complete from the above inequality.
\end{IEEEproof}
 As a result of the above lemma and \eqref{eq: bound on log M_i for MAC Exp}, the inequality  $\EE[\tau_i]\geq \frac{\log M_i}{I_L^i+\epsilon}-\frac{h_b(\delta)}{I_L^i+\epsilon}$ holds. Finally, combining this inequality with \eqref{eq: P_e^1}-\eqref{eq: P_e^3} completes the proof of the theorem.

\subsection{An Alternative Proof for the Upper-Bound}\label{subsec: proof of theorem Burnashev}

In this part of the paper, we provide a series of Lemmas that are used to prove the Theorem. 
 Define the following random processes.
%

\begin{lem}\label{lem: generalized fano for MAC-FB}
For an $(M_1, M_2, N)$-VLC with probability of error $P_e$ the following inequality holds
\begin{align*}
\EE[H^i_T]\leq h_b(P_e)+P_e \log_2(M_1M_2-1), \quad \text{for} \quad  i=1,2,3.
\end{align*}
\end{lem}

\begin{IEEEproof}
The proof follows from Fano's Lemma as in \cite{Burnashev}.
\end{IEEEproof}
%


\begin{lem}\label{lem: log drift MAC-FB}
There exists $\epsilon>0$ such that, if $H_t^i \leq \epsilon$, then 
\begin{align*}
\EE[\log H_{t+1}^1-\log H^1_{t}|\mathcal{F}_{W_2}\times\mathcal{F}_t] &\geq -(D_1+\epsilon),\\
\EE[\log H_{t+1}^2-\log H^2_{t}|\mathcal{F}_{W_1}\times \mathcal{F}_t] &\geq -(D_2+\epsilon),\\
\EE[\log H_{t+1}^3-\log H^3_{t}|\mathcal{F}_t] &\geq -(D_3+\epsilon)
\end{align*}
holds almost surely, where $D_i, i=1,2,3$ are defined in \eqref{eq: definiton of D_i}.
\end{lem}
\begin{IEEEproof}
The proof is given in Appendix \ref{appx: proof of lem log drift}.
\end{IEEEproof}

%

\begin{lem}\label{lem: Z_i are submartingale}
For $i=1,2,3$, define  random process $\{Z^{(i)}_t\}_{t\geq 1}$ as 
\begin{align}\nonumber
Z^{(i)}_t =&\left(\frac{\log H^i_t-\log \epsilon}{D_i}+t+f_i(\log \frac{H^i_t}{\epsilon})\right)\mathbbm{1}{\{H^i_t \leq \epsilon\}}\\\label{eq: Z_t^i}
&+ \left(\frac{H^i_t-\epsilon}{I^i_L}+t\right)\mathbbm{1}{\{H^i_t \geq \epsilon	\}}
\end{align} 
where the function $f_i$ is defined as $f_i(y)=\frac{1-e^{-\mu_i y}}{D_i \mu_i}.$ Then, there exists $\mu_i>0$ such that $Z^{(i)}_t$ is a submartingale w.r.t $\mathcal{F}_{W_1}\times \mathcal{F}_{W_2}\times\mathcal{F}_t$.
\end{lem}
\begin{IEEEproof}[Outline of the proof]
Suppose $W_2=m$ for some $m\in [1:M_2]$. Given this event and using the same argument as in the proof of Theorem 1 in \cite{Burnashev} we can show that $Z_{t}^{(i)}| W_2=m$ is a submartingale for all $m$. More precisely, the inequality
\begin{align*}
\EE\{Z_{t}^{(i)}-Z_{t+1}^{(i)}| \mathcal{F}_{W_1}\times \mathcal{F}_{W_2}\}\leq 0,
\end{align*} 
holds almost surely w.r.t $\mathcal{F}_{W_1}\times \mathcal{F}_{W_2}$. Taking the expectation of the both sides in the above inequality gives
\begin{align*}
\EE\{Z_{t}^{(i)}-Z_{t+1}^{(i)}\}\leq 0,\quad \forall t\geq 0, ~ i=1,2,3.
\end{align*}
Thus, $Z_{t}^{(i)}$ is a submartingale for $i=1,2,3$ and w.r.t $\mathcal{F}_{W_1}\times \mathcal{F}_{W_2}\times \mathcal{F}_t$.
\end{IEEEproof}
\begin{corollary}
Suppose  $\alpha_1, \alpha_2, \alpha_3$ are non-negative numbers such that $\alpha_1+\alpha_2+\alpha_3=1$. Define $Z_t = \alpha_1 Z_t^{(1)}+\alpha_2 Z_t^{(2)}+\alpha_3 Z_t^{(3)}$. Then, $Z_t$ is a submartingale w.r.t $\mathcal{F}_{W_1}\times \mathcal{F}_{W_2}\times\mathcal{F}_t$.  
\end{corollary}

The Theorem follows from the above lemma, and the proof is given in Appendix \ref{appx: proof of thm Error Exp Upper bound}.

\section{The Shape of the Lower and Upper Bounds}\label{sec: the shape of the bounds}
In this Section, we point out a few remarks on $E_u(R_1,R_2)$ and the lower-bound $E_l(R_1,R_2)$ defined in Theorem \ref{thm: Error Exp lowerbound}. Furthermore, we provide an alternative representation for the bounds and show that the lower and upper-bounds match for a class of MACs.

We first compare the lower bound in \eqref{eq: E_l} and the upper-bound in Corollary \ref{cor: Exp upper bound}. For a given arbitrary rate pair $(R_1,R_2)$ inside the feedback-capacity of a given MAC, consider a sequence of VLCs with rates $(R_1,R_2)$ and with average probability of error approaching zero. Then, the following holds:
$$\lim_{\epsilon\rightarrow 0}\lim_{P_e \rightarrow 0}\lim_{M_1M_2 \rightarrow \infty}\frac{E_u(R_1,R_2)}{E_l(R_1,R_2)}=\frac{D_u}{D_l}$$  
As a result of the above remark, it is concluded that for small enough probability of error, the bounds are different only in the constants $D_u$ and $D_l$.

Next, provide an alternative representation for the lower/upper-bound. For that, suppose $(R_1,R_2)$ is a point inside the capacity region $\mathcal{C}$. By $(||\underline{R}||, \theta_R)$ denote the polar coordinate of $(R_1,R_2)$ in $\RR^2$. It is shown in the following Remark that the optimum $\underline{\lambda}$ in $E_u$ and $E_l$ is independent of the Euclidean norm of $(R_1,R_2)$, i.e., $\|\underline{R}\|$. 
\begin{remark}
Given an arbitrary $\alpha>0$ and a rate pair $(R_1,R_2)$ in the capacity region, the optimum $\underline{\lambda}$ for $E_l(R_1,R_2)$ is the same as the one for $E_l(\alpha R_1, \alpha R_2)$.
\end{remark}
\begin{IEEEproof}
Note that one can write $E_l(R_1,R_2)$ as 
\begin{align*}
E_l(R_1,R_2)&= D_l\left( 1- \max_{\substack{\lambda_1, \lambda_2, \lambda_3 \geq 0\\ \lambda_1+\lambda_2+\lambda_3=1}} \frac{\sum_{i=1}^3  \lambda_i R_i}{C_{\lambda}}\right),\\
&= D_l\left( 1- \frac{\sum_{i=1}^3  \lambda^*_i R_i}{C_{\lambda^*}}\right),
\end{align*}
where $\underline{\lambda}^* $ is the optimum $\underline{\lambda}$ for $E_l$. Next, replace $(R_1, R_2)$ with $(\alpha R_1, \alpha R_2)$ for some constant $\alpha>0$. Then, we obtain 
\begin{align*}
E_l(\alpha R_1, \alpha R_2)&= D_l \left( 1- \alpha \max_{\substack{\lambda_1, \lambda_2, \lambda_3 \geq 0\\ \lambda_1+\lambda_2+\lambda_3=1}} \frac{\sum_{i=1}^3  \lambda_i R_i}{C_{\lambda}}\right),\\
&\stackrel{(a)}{=} D_l\left( 1- \alpha \frac{\sum_{i=1}^3  \lambda^*_i R_i}{C_{\lambda^*}}\right),
\end{align*}
where (a) follows as the objective function for the maximization is the same as the one in $E_l(R_1,R_2)$. This implies that there is an identical $\underline{\lambda}^*$ which optimizes the  expression in $E_l(R_1,R_2)$ and $E_l(\alpha R_1, \alpha R_2)$. 
\end{IEEEproof}
\begin{figure}[hbtp]
\centering
\includegraphics[scale=0.7]{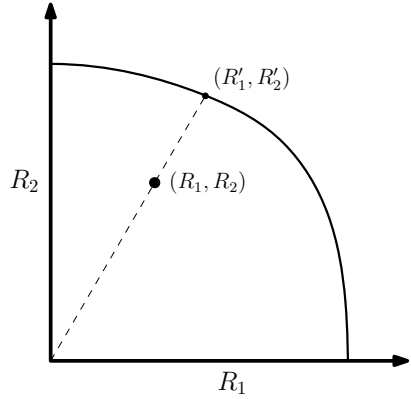}
\caption{Given a rate pair $(R_1, R_2)$ which is inside the capacity region, consider the line passing $(R_1,R_2)$ and the origin. Then, $(R'_1, R'_2)$ is the point of intersection of this line with the boundary of the capacity region. }
\label{fig: R' and R}
\end{figure}

\begin{figure}[hbtp]
\centering
\includegraphics[scale=0.4]{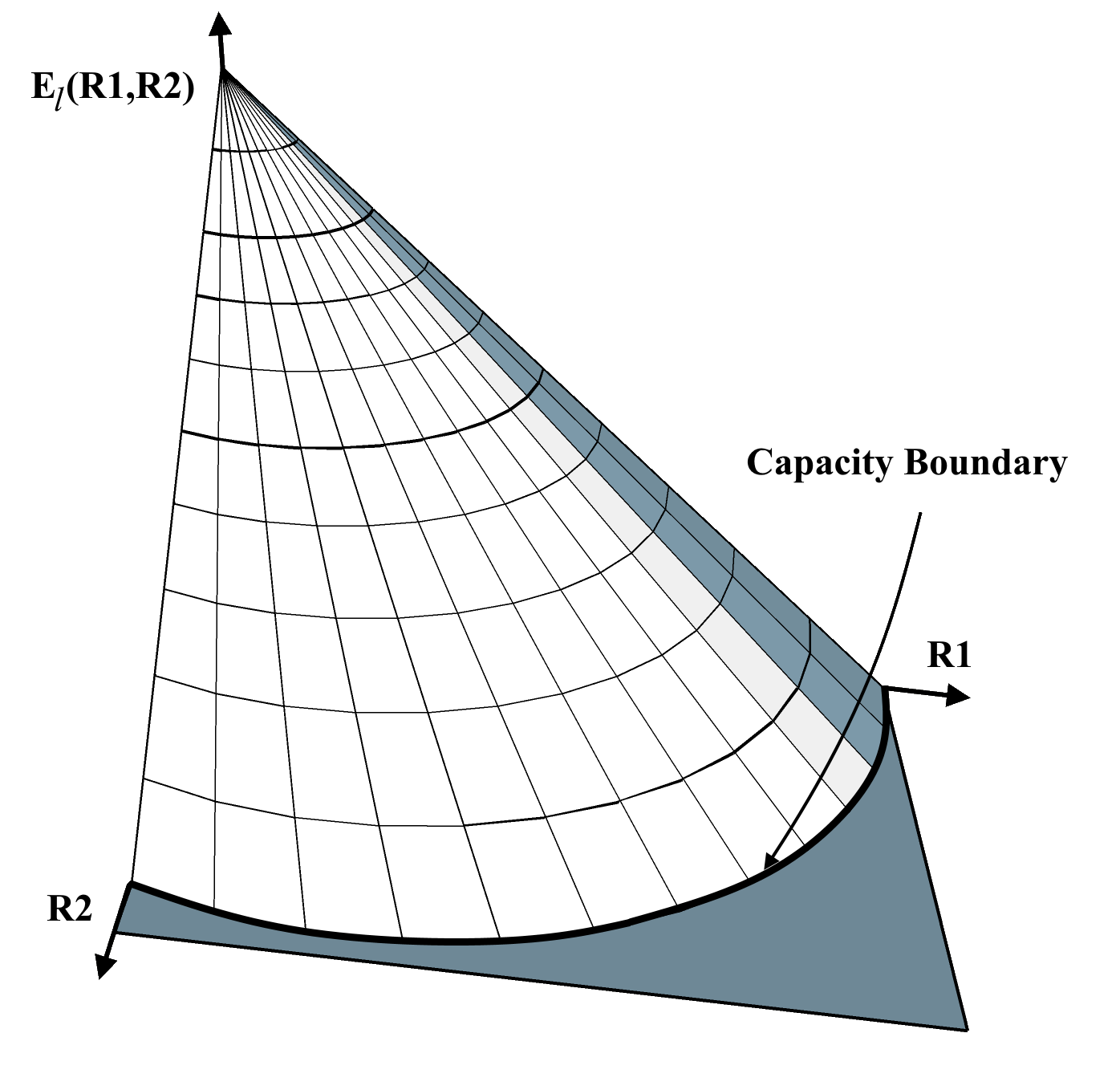}
\caption{The conceptual shape of the lower/upper-bound on the error exponent of a given MAC with respect to the transmission rate pair $(R_1,R_2)$.}
\label{fig: shape of the bounds}
\end{figure}

Now, consider the line passing $(R_1,R_2)$ and the origin. Let $(R'_1, R'_2)$ denote the point of intersection of this line with the boundary of the capacity region. Fig. \ref{fig: R' and R} shows how $(R'_1, R'_2)$ is determined.  Since, $R'_i=\alpha R_i,i=1,2$ for some $\alpha>0$, then the optimum $\underline{\lambda}$ in $E_l(R'_1, R'_2)$ is the same as the one in $E_l(R_1, R_2)$. Therefore, from this argument and the fact that $R_i=\frac{R'_i}{\alpha}, i=1,2$, we can rewrite $E_l(R_1,R_2)$ as 
\begin{align*}
E_l(R_1,R_2)&= \min_{\substack{\lambda_1, \lambda_2, \lambda_3 \geq 0\\ \lambda_1+\lambda_2+\lambda_3=1}}D_{l}\left(1-\frac{1}{\alpha} \frac{\sum_{i=1}^3  \lambda_i R'_i}{C_{\lambda}}\right),\\
&\stackrel{(a)}{=}D_{l}\left(1- \frac{1}{\alpha}\right),
\end{align*}
where $(a)$ follows, since $(R'_1, R'_2)$ is on the capacity boundary. Note that $\alpha=\frac{\|\underline{R}\|}{\|\underline{R'}\|}$. Therefore, $E_l(R_1,R_2)=D_{l}\left(1- \frac{\|\underline{R}\|}{\|\underline{R'}\|}\right)$. Moreover, note that $\|\underline{R'}\|$ depends on $(R_1,R_2)$ only through $\theta_R$; in particular, it equals to $C(\theta_R)$ which is a function of $\theta_R$. With this notation, we can rewrite $E_l$ as
$$E_l(R_1,R_2)=D_{l}\left(1- \frac{\|\underline{R}\|}{C(\theta_R)}\right)$$
Using a similar argument for $E_u$, we have 
$$E_u(R_1,R_2)=D_{u}\left(1- \frac{\|\underline{R}\|}{C(\theta_R)}\right)+\delta.$$
As a conclusion of the above argument, the lower (upper) bound increases linearly with respect to a specific Euclidean distance measure defined between the transmission rate pair and the capacity boundary. Fig. \ref{fig: shape of the bounds}  shows the shape of a typical upper (lower) bound as a function of the transmission rate pairs.

\subsection{On the Tightness of the Bounds on the Error Exponent }
In what follows, we provide examples of classes of channels for which the lower and upper bound coincide.

\begin{example}
Consider a MAC in which the output is $(Y_1,Y_2)$ and the transition probability matrix is described by the product $Q_{Y_1|X_1}Q_{Y_2|X_2}$. This MAC consists of two parallel (independent) point-to-point channels. Suppose, $C_1$ and $C_2$ are the capacity of the first and the second parallel channel, respectively. For this MAC, one can use two parallel Yamamoto-Itoh schemes, one for each channel. Based on the results for the point-to-point case, it is not difficult to show that the error exponent  for such MAC satisfies 
\begin{align}\label{eq: ex lowerbound}
E(R_1,R_2)\geq \min\{D_1(1-\frac{R_1}{C_1}), D_2(1-\frac{R_2}{C_2}) \},
\end{align}
where $C_1$ and $C_2$ are the point-to-point capacity of the channel corresponding to $Q_{Y_1|X_1}$ and $Q_{Y_2|X_2}$, respectively. Note that this lower-bound is not covered by the proposed coding strategy given  in Section \ref{sec: lower bound}. 
For such MAC, the upper-bound given in \eqref{eq: E_u} is simplified to 
\begin{align*}
E(R_1,R_2)&\leq \min_{\substack{ \lambda_1, \lambda_2 \geq 0  }}\min_{j\in \{1,2\}} D_j \left(1- \frac{\lambda_j R_j}{\lambda_1 C_1+\lambda_2 C_2}\right)+\delta.
\end{align*}
The right-hand side of the above inequality is further upper-bounded by substituting $(\lambda_1, \lambda_2)=(0,1)$ or $(\lambda_1, \lambda_2)=(1,0)$. Therefore, we obtain
\begin{align*}
E(R_1,R_2)&\leq\min_{j\in \{1,2\}} D_j \left(1- \frac{R_j}{C_j}\right)+\delta
\end{align*}
By letting $\delta\rightarrow 0$ as in Theorem \ref{thm: Error Exp Upper bound}, the above bound can be made arbitrary close to the lower-bound given in \eqref{eq: ex lowerbound}.

\end{example}

\begin{example}
Consider a MAC with input alphabets $\mathcal{X}_1=\mathcal{X}_2=\{0,1,2\}$, and output alphabet $\mathcal{Y}=\{0,1,2\}$. The transition probability of the channel is described by the following relation:
$$Y=X_1 \oplus _3 X_2 \oplus_3 N_p,$$
where the additions are modulo-3 addition, and $N_p$ is a random variable with $P(N_p=1)=P(N_p=2)=p$, and $P(N_p=0)=1-2p$, where $0\leq p\leq 1/2$. 
It can be shown that for this channel $D_l=D_u=(1-3p) \log \frac{1-2p}{p}.$
Hence, the upper-bound in Corollary \ref{cor: Exp upper bound} can be made arbitrary close to the lower-bound in Theorem \ref{thm: Error Exp lowerbound}.

\end{example}
The argument in the above example can be extended to $m$-ary additive MACs for $m>2$, where the transition probability of the channel is described by 
$$Y=X_1 \oplus_m X_2 \oplus_m N_p,$$
where all the random variables take values from $\ZZ_m$, and $N_p$ is a random variable with $P(N_p=i)=p$ for any $i\in \ZZ_m, ~i\neq 0$ and $P(N_p=0)=1-(m-1)p$. It can be shown that for this channel 
$$D_l=D_u=(1-mp) \log \frac{1-(m-1)p}{p}.$$


%
%

\section{Conclusion }\label{sec: conclution}
We derive a lower and upper-bound on the reliability function of discrete memoryless MAC with noiseless feedback and variable-length codes. For the lower-bound, we adapt Yamamoto and Itoh's coding scheme consisting of a data and a confirmation phase. For the upper-bound, we adopt the proof techniques of Burnashev for the reliability function of the point-to-point case. The two bounds have the same shape with the difference being the constants at zero rate. 
We identify sequential binary hypothesis testing problems that are used to derive the upper-bound. 
We show that the bounds are tight for a class of MACs. 
\appendices

\section{Proof of Theorem \ref{thm: Error Exp lowerbound} }\label{appx: proof of thm error Exp lowebound}
\begin{IEEEproof}
{
At each block a re-transmission occurs  with probability $q$, an error occurs with probability $P_{eb}$ and a correct decoding process happens with probability $1-q-P_{eb}$. 
The probability of a re-transmission at each block is 
\begin{align*}
q=P(\hat{\Theta}=\Theta_1).
\end{align*}
The probability of error at each block is
\begin{align*}
P_{eb}=P(\Theta_1)P(\hat{\Theta}=\Theta_0|\Theta_1).
\end{align*}
 Therefore, with this setting the total probability of error for the transmission of a message is 
\begin{align}\label{eq: P_e in terms of P_eb}
P_e=\sum_{k=0}^{\infty} q^k P_{eb}=\frac{P_{eb}}{1-q}.
\end{align}
 
The number of blocks required to complete the transmission of one message is a geometric random variable with probability of success $1-q$. Thus, the expected number of blocks for transmission of a message is $\frac{1}{1-q}$. 
}

Next, we derive an upper-bound for $q$ and $P_{eb}$. 
For shorthand,  denote $H_{12}=(H_1, H_2), \hat{H}_{12}=(\hat{H}_1, \hat{H}_2)$. Then
\begin{align*}
P_{eb}&=P\left(\hat{H}_{12}=00, H_{12}\neq 00\right)\\
&=\sum_{a\in \{01,10,11\}}P(H_{12}=a)P(\hat{H}_{12}=00|H_{12}=a).
\end{align*}

Note that the effective rates of this transmission scheme are $(\frac{R_1}{1-\gamma}, \frac{R_2}{1-\gamma})$. Suppose $(\frac{R_1}{1-\gamma}, \frac{R_2}{1-\gamma})$ is inside the feedback-capacity region of the channel. Then, from the definition of the capacity region, there exist a sequence $\zeta_n , n\geq 1$ with $\zeta_n \rightarrow 0$ such that after the first stage $$P((\hat{W}_1,\hat{W}_2)\neq (W_1,W_2))\leq \zeta_n.$$ Equivalently, the effective rates are inside the capacity region, if the following inequality holds for any $\lambda_i \geq 0, i=1,2,3$:
\begin{align}\label{eq: condition to be inside capacity}
\frac{1}{1-\gamma} \left(\lambda_1 R_1 +\lambda_2 R_2 +\lambda_3 (R_1+R_2) \right)< C_{\underline{\lambda}},
\end{align}
where $C_{\underline{\lambda}}$ is given in Definition \ref{def: C lambda}. Denote $R_3 =R_1+R_2$ and define 
\begin{align}\label{eq: gamma^*}
\gamma^*= \min_{\substack{\lambda_1, \lambda_2, \lambda_3 \geq 0\\ \lambda_1+\lambda_2+\lambda_3=1}} (1-\frac{\sum_i \lambda_i R_i}{C_{\underline{\lambda}}}).
\end{align}
Then, \eqref{eq: condition to be inside capacity} implies that $\gamma < \gamma^*$. 
The probability of error is therefore bounded by
\begin{align} \label{eq: bound ofr P_e for confirmation stage}
P_{eb}&\leq  \sum_{a\in \{01,10,11\}}P(\hat{H}_{12}=00|H_{12}=a)
\end{align}

%
Suppose $(X_1(0), X_1(1), X_2(0),X_2(1))$ are random variables with joint distribution $\mathsf{P}_n$. Then for $i,j \in \{0,1\}$ define
\begin{align*}
\bar{D}_{\mathsf{P}_n}(00||ij)=\EE_{\mathsf{P}_n}\Big[D_Q\big(X_1(0), X_2(0)|| X_1(i), X_2(j)\big)\Big].
\end{align*}
From the description of the transmission scheme, the codewords for the confirmation stage are selected with joint-type $\mathsf{P}_n$. In addition, the decoding process is performed using ML decoding. Therefore, the following bounds hold for $a\in \{01,10,11\}$:   
\begin{align*}
P(\hat{H}_{12}=00|H_{12}=a) \leq 2^{-n\gamma \bar{D}_{\mathsf{P}_n}(00||a)}.
\end{align*}
Thus, from \eqref{eq: bound ofr P_e for confirmation stage}, the probability of error is upper bounded by 
\begin{align}\label{eq: bound on Pe}
P_{eb}&\leq 3 \times  2^{-n\gamma D_{l,n}}
\end{align} 
Where $D_{l,n}= \max_{\mathsf{P}_n}  \min_{a\in \{01,10,11\}} \bar{D}_{\mathsf{P}_n}(00||a)$. 

Next we derive an upper bound for $q$. We have 
\begin{align*}
q&=P(\hat{\Theta}=\Theta_1)\\
&=P(\Theta_0)P(\hat{\Theta}=\Theta_1|\Theta_0)+P(\Theta_1)P(\hat{\Theta}=\Theta_1|\Theta_1)\\
&\leq P(\hat{\Theta}=\Theta_1|\Theta_0)+\zeta_n,
\end{align*}
where the last inequality holds  because of the following inequalities  1) $P(\Theta_1)\leq \zeta_n$, and 2) $P(\Theta_0), P(\hat{\Theta}=\Theta_1|\Theta_1) \leq 1$. Note that 
\begin{align*}
 P(\hat{\Theta}=\Theta_1|\Theta_0)&=\sum_{a\in \{01,10,11\}} P(\hat{H}_{12}=a|H_{12}=00)\\
 &\leq \sum_{a\in \{01,10,11\}} 2^{-n\gamma \bar{D}_{\mathsf{P}_n}(a||00)}\\
 &\leq 3 \times 2^{-n\gamma \tilde{D}_{l,n}},
\end{align*}
where $\tilde{D}_{l,n}= \min_{a\in \{01,10,11\}} \bar{D}_{\mathsf{P}_n}(a||00)$. 
Therefore, there exists a sequence $\{q_n\}_{n\geq 1}$ with $q_n \rightarrow 0$ such that 
$q < q_n+\zeta_n$. Using this inequality and the inequality at \eqref{eq: bound on Pe}, we derive the following upper-bound for the total probability of error given in \eqref{eq: P_e in terms of P_eb} 
\begin{align*}
P_e \leq \frac{3 }{1-q_n-\zeta_n}  2^{-n\gamma D_{l,n}}.
\end{align*}
 Therefore, the error exponent is bounded from below as 
\begin{align*}
\frac{-\log_2 P_e}{\EE[T]} &\geq \sup \frac{\gamma D_{l,n}}{(1-q_n-\zeta_n)}+\xi_n
\end{align*}
where $\xi_n=\frac{1}{n}\frac{\log_2(\frac{1-q_n-\zeta_n}{3})}{1-q_n-\zeta_n}$. Note that for any $\epsilon>0$ there exists large enough $n$ such that $q_n+\zeta_n<\epsilon, D_{l,n}> D_l-\epsilon, \xi_n <\epsilon$. Set $\gamma=\gamma^*-\epsilon$. Then
\begin{align*}
\frac{-\log_2 P_e}{\EE[T]} &\geq \gamma^* D_{l}-\sigma(\epsilon)
\end{align*}
where $\sigma$ is a function of $\epsilon$ such that $\lim_{\epsilon \rightarrow 0}\sigma(\epsilon)=0$. Finally, the proof is complete by replacing $\gamma^*$ from \eqref{eq: gamma^*}.

\end{IEEEproof}

{
\section{Proof of Lemma \ref{lem: linear drift MAC FB}}\label{appx: proof of lem linear drift}
\begin{IEEEproof}
Given $Y^t=y^t, W_1=m_1, W_2=m_2$, we obtain 
\begin{align}\nonumber
\EE[& H^1_{t+1}-H^1_{t}|m_2, y^t ]\\\nonumber
&= -I(W_1;Y_{t+1}|m_2, y^t )\\\nonumber
&= -I(W_1;Y_{t+1}|m_2,  x_{2}^{t+1}, y^t )\\\nonumber
&=-H(Y_{t+1}|m_2, x_{2}^{t+1}, y^t )+H(Y_{t+1}|m_2, x_{2}^{t+1},W_1, y^t)\\ \nonumber
&=-H(Y_{t+1}|m_2, x_{2}^{t+1}, y^t )\\\nonumber
& \quad +H(Y_{t+1}|m_2, x_{2}^{t+1},W_1, X_{1}^{t+1}, y^t)\\\nonumber
&\stackrel{(a)}{=}-H(Y_{t+1}|m_2, x_{2}^{t+1}, y^t )  +H(Y_{t+1}|x_{2}^{t+1}, X_{1}^{t+1},y^t)
\label{eq: bound on linear drift of H^1 }\\
&\triangleq -J^1_{t+1}(m_2,x_{2}^{t+1},y^t)
\end{align}
where $(a)$ follows because condition on the channel inputs $X_{1, t+1}, X_{2, t+1}$, the output $Y_{t+1}$ is independent of $W_1,W_2$. We denote the right-hand side of $(a)$ by $J^1_{t+1}(.)$ as in \eqref{eq: bound on linear drift of H^1 }. 
Similarly for the case when $i=2$ the following lower-bound holds 
\begin{align}\nonumber
\EE[&H^2_{t+1}-H^2_{t}|m_1,y^t ]\\\nonumber
&= -H(Y_{t+1}|m_1, x_{1}^{t+1}, y^t )  +H(Y_{t+1}|X_{2}^{t+1}, x_{1}^{t+1},y^t)\\\label{eq: bound on linear drift of H^2 }
&\triangleq -J^2_{t+1}(m_1,x_{1}^{t+1},y^t).
\end{align}
Using a similar argument for the case when $i=3$, we can show that the following inequality holds 
\begin{align}\nonumber
\EE[H^3_{t+1}-H^3_{t}|y^t ]&\geq -I(X_{1}^{t+1},X_{2}^{t+1};Y_{t+1}| y^t)\\\label{eq: bound on linear drift of H^3 }
&\triangleq -J^3_{t+1}(y^t).
\end{align}
Consider the quantities at the right-hand side of \eqref{eq: bound on linear drift of H^1 }, \eqref{eq: bound on linear drift of H^2 } and \eqref{eq: bound on linear drift of H^3 }, i.e., the functions $J^1_{t+1}, J^2_{t+1},J^3_{t+1}$. We proceed by the following lemma. 

\begin{lem}
The vector $(J^1_{t+1}, J^2_{t+1},J^3_{t+1})$ is inside the feedback-capacity region $\mathcal{C}$ almost surely.
\end{lem}
\begin{IEEEproof}
We use the alternative representation for $\mathcal{C}$ which is given in Fact \ref{Fact: MAC-FB capacity hyperplane}. For any non-negative numbers $\lambda_1, \lambda_2, \lambda_3$, let 
\begin{align*}
 J_{\lambda}(m_1,m_2,x_1^{t+1}, x_2^{t+1}, y^t)&=\lambda_1 J^1_{t+1}(m_2,x_{2}^{t+1},y^t)+\lambda_2 J^2_{t+1}(m_1,x_{1}^{t+1},y^t)+\lambda_3 J^3_{t+1}(y^t)
\end{align*}
 Note that 
\begin{align}\label{eq: bound on J lambda}
J_{\lambda}(m_1,& m_2, x_1^{t+1}, x_2^{t+1},y^t) \leq \sup_{P_{W_1W_2X^{t+1}_1X^{t+1}_2|Y^{t+1}}} \EE\{J_{\lambda}(W_1,W_2,X_1^{t+1}, X_2^{t+1}, y^t)\},
\end{align}
where the supremum is taken over all $P_{X^{t+1}_1X^{t+1}_2|Y^{t+1}}$ that factors as in \eqref{eq: factorization of P}.
The right-hand side of the above inequality  equals $\sum_i \EE[\lambda_i J^i_{t+1}]$. Each expectation inside the summation can be bounded as follows  
\begin{align*}
\EE\{J^1_{t+1}(W_2,X_{2}^{t+1},y^t)\}&=H(Y_{t+1}|W_2, X_{2}^{t+1}, y^t ) - H(Y_{t+1}|X_{2}^{t+1}, X_{1}^{t+1},y^t)\\
&\leq H(Y_{t+1}|X_{2}^{t+1}, y^t ) - H(Y_{t+1}|X_{2}^{t+1}, X_{1}^{t+1},y^t)\\
&=I(X_{1, t+1}, Y_{t+1}|X_{2}^{t+1}, y^t)
\end{align*}
Similarly, 
\begin{align*}
\EE\{J^2_{t+1}(W_1,X_{1}^{t+1},y^t)\}&\leq I(X_{2, t+1}; Y_{t+1}|X_1^{t+1}y^t)\\
\EE\{J^3_{t+1}(y^t)\}&\leq I(X_{1,t+1},X_{2,t+1};Y_{t+1}| y^t)
\end{align*}
Therefore, since the channel is memoryless using the above bounds we have
\begin{align*}
&\EE\{J_{\lambda}(W_1,W_2,X_1^{t+1}, X_2^{t+1}, y^t)\}\\
&\leq\lambda_1 I(X_{1, t+1}, Y_{t+1}|X_{2}^{t+1}, y^t) +\lambda_2 I(X_{2, t+1}; Y_{t+1}|X_1^{t+1}y^t)+\lambda_3 I(X_{1,t+1},X_{2,t+1};Y_{t+1}| y^t)\\
&\leq C_{\underline{\lambda}}
\end{align*}
\end{IEEEproof}
Since the vector $(J^1_{t+1}, J^2_{t+1},J^3_{t+1})$ is inside the capacity for all $1\leq t \leq N$, then, by definition, $\forall \epsilon>0$ there exist $L$ and $P_{X^{L}_1X^{L}_2 Y^{L}}$ factoring as in \eqref{eq: factorization of P} such that
 \begin{align*}
 J^i_{t+1} \leq I^i_L+\epsilon, \quad i=1,2,3
  \end{align*}
holds for all $1\leq t \leq N$. This implies the statement of the lemma.

\end{IEEEproof}
}

{
\section{Proof of Lemma \ref{lem: log drift MAC-FB} }\label{appx: proof of lem log drift}
\begin{IEEEproof}
We prove the first statement of the lemma. The second and the third statements follow by a similar argument. Given $Y^t=y^t, W_2=m$, define the following quantities
\begin{align*}
f_{i|m}&=P(W_1=i| Y^t=y^t, W_2=m)\\
f_{i|m}(y_{t+1})&=P(W_1=i| Y^t=y^t, W_2=m, Y_{t+1}=y_{t+1})\\
Q_{i,m}(y_{t+1})&=P(Y_{t+1}=y_{t+1}|W_1=i, W_2=m, Y^t=y^t),
\end{align*}
where $i\in [1:M_1], y_{t+1}\in \mathcal{Y}$.
Since $H^1_t < \epsilon$, then there exist $\epsilon'$ (as a function of $\epsilon$) and an index $l\in [1:M_1]$ such that $f_{l|m}\geq 1-\epsilon'$ and $f_{i|m}\leq \frac{\epsilon'}{M_1-1}$ for all $i\in [1:M_1], i\neq l$. Denote
\begin{align*}
\hat{f}_{i|m}=\frac{f_{i|m}}{1-f_{l|m}}, \quad  i\neq l.
\end{align*}
Using the grouping axiom we have
\begin{align*}
H^1_{t}=H(W_1|W_2=m, y^t)=h_b(f_{l|m})+(1-f_{l|m})H(\hat{X})
\end{align*}
where $\hat{X}$ is a random variable with probability distribution $P(\hat{X}=i)=\hat{f}_{i|m}, ~i\in [1:M_1], i\neq l$. Note that $$h_b(f_{l|m}) \approx -(1-f_{l|m})\log (1-f_{l|m}).$$ Therefore,
\begin{align}\nonumber
H^1_{t}&\approx -(1-f_{l|m})(\log(1-f_{l|m})-H(\hat{X}))\\\label{eq: approximate H^1_t}
&\approx (1-f_{l|m})\log(1-f_{l|m})
\end{align}
where the last approximation is due to the fact that $-\log(1-f_{l|m})\gg H(\hat{X})$. 
 Next, we derive an approximation for $H^1_{t+1}$. Note that
\begin{align*}
f_{l|m}(y_{t+1})=\frac{f_{l|m}Q_{l,m}(y_{t+1})}{\sum_j f_{j|m}Q_{j,m}(y_{t+1})}
\end{align*}
The denominator can be written as 
$$f_{l|m}Q_{l,m}(y_{t+1})+(1-f_{l|m})\sum_{j\neq l} \hat{f}_{j|m}Q_{j,m}(y_{t+1}).$$
The above quantity is approximately equals to $Q_{l,m}(y)$. Therefore,
\begin{align*}
(1-f_{l|m}(y_{t+1}))&=(1-f_{l|m}) \frac{\sum_{j\neq l} \hat{f}_{j|m}Q_{j,m}(y_{t+1})}{\sum_j f_{j|m}Q_{j,m}(y_{t+1})}\\
&\approx (1-f_{l|m}) \frac{\sum_{j\neq l} \hat{f}_{j|m}Q_{j,m}(y_{t+1})}{Q_{l,m}(y_{t+1})}
\end{align*}
This implies that $f_{l|m}(y_{t+1}) \approx 1$. Therefore, using the same argument for $H^1_t$ we have
\begin{align}\nonumber
H^1_{t+1}&\approx -(1-f_{l|m}(y_{t+1}))(\log(1-f_{l|m}(y_{t+1}))\\\nonumber
&=-(1-f_{l|m}(y_{t+1}))\Big[\log(1-f_{l|m})+ \log (\frac{\sum_{j\neq l} \hat{f}_{j|m}Q_{j,m}(y_{t+1})}{Q_{l,m}(y_{t+1})})\Big]\\\label{eq: approximate H^1_t+1}
&\approx -(1-f_{l|m}(y_{t+1}))\log(1-f_{l|m}).
\end{align}
As a result of the approximations in \eqref{eq: approximate H^1_t} and \eqref{eq: approximate H^1_t+1}, we obtain
\begin{align*}
\frac{H^1_{t+1}}{H^1_{t}}&\approx \frac{(1-f_{l|m}(y))\log(1-f_{l|m})}{(1-f_{l|m})\log(1-f_{l|m})}\\
&=\frac{\sum_{j\neq l} \hat{f}_{j|m}Q_{j,m}(y)}{Q_{l,m}(y)}
\end{align*}
Note that
\begin{align*}
P(Y_{t+1}=y| W_2=m, y^t)\approx Q_{l,m}(y)
\end{align*}
Therefore, 
\begin{align*}
\EE\{\log \frac{H^1_{t+1}}{H^1_{t}}| y^t\}&\approx \EE\{\log \frac{\sum_{j\neq l} \hat{f}_{j|m}Q_{j,m}(Y_{t+1})}{Q_{l,m}(Y_{t+1})}\}\\
&=\sum_{y}Q_{l,m}(y) \log \frac{\sum_{j\neq l} \hat{f}_{j|m}Q_{j,m}(y)}{Q_{l,m}(y)}\\
&\stackrel{(a)}{=}-D(Q_{l,m}|| \sum_{j\neq l} \hat{f}_{j|m}Q_{j,m})\\
& \stackrel{(b)}{\geq}  -\sum_{j\neq l} \hat{f}_{j|m} D(Q_{l,m}|| Q_{j,m})\\
&\geq -\max_{j\neq l} D(Q_{l,m}|| Q_{j,m})\\
&\stackrel{(c)}{\geq}  -(D_1+\epsilon)
\end{align*}
where $(a)$ is due to the definition of Kullback–Leibler divergence, $(b)$ is due to the convexity of Kullback–Leibler divergence, and $(c)$ is due to the definition of $D_1$.
\end{IEEEproof}

}


%
\section{Proof of Theorem \ref{thm: Error Exp Upper bound}}\label{appx: proof of thm Error Exp Upper bound}

\begin{proof}
Since $\{Z_t\}$ is a submartingale, then $Z_0 \leq \EE[Z_T]$. By the definition of $\{Z_t\}$  we have $\EE[Z_T]=\sum_{i=1}^3\alpha_i \EE[Z_T^i].$ For any of processes $\{Z^i_t\}$, the following hold:
\begin{align}\nonumber
\EE[Z^i_T]&=\EE \left[ \frac{H^i_T-\epsilon}{I^i_L+\epsilon}1_{\{H^i_T \geq \epsilon	\}}\right]+ \EE \left[ \left(\frac{\log H^i_T-\log \epsilon}{D_i+\epsilon}+f_i(\log \frac{H^i_T}{\epsilon})\right)1_{\{H^i_T \leq \epsilon\}}\right]+\EE[T]\\\nonumber
&\leq \EE \left[ \frac{H^i_T-\epsilon}{I^i_L+\epsilon}\right]+ \EE \left[\frac{\log H^i_T-\log \epsilon}{D_i+\epsilon}+f_i(\log \frac{H^i_T}{\epsilon})\right]+\EE[T]\\\nonumber
&\stackrel{(a)}{\leq} \EE \left[ \frac{H^i_T-\epsilon}{I^i_L+\epsilon}\right]+ \EE \left[\frac{\log H^i_T-\log \epsilon}{D_i+\epsilon}\right]+\frac{1}{\mu_i D_i}+\EE[T]\\\nonumber
& = \frac{\EE[H^i_T]-\epsilon}{I^i_L+\epsilon}+ \frac{\EE[\log H^i_T]-\log \epsilon}{D_i+\epsilon}+\frac{1}{\mu_i D_i}+\EE[T]\\\label{eq: bound on E[Z_T^i]}
&\stackrel{(b)}{\leq}  \frac{\EE[H^i_T]-\epsilon}{I^i_L+\epsilon}+ \frac{\log \EE[H^i_T]-\log \epsilon}{D_i+\epsilon}+\frac{1}{\mu_i D_i}+\EE[T]
\end{align}
where $(a)$ follows from the inequality $f_i(y)\leq \frac{1}{\mu_i D_i} $, and $(b)$ follows by applying Jensen's inequality for the function $\log(x)$.

Define $\eta(P_e)=h_b(P_e) + P_e \log(M_1M_2)$. Using Lemma \ref{lem: generalized fano for MAC-FB}, the right-hand side of \eqref{eq: bound on E[Z_T^i]} is upper bounded as 
\begin{align}\nonumber
&\leq \frac{\eta(P_e)-\epsilon}{(I^i_L+\epsilon)}+ \frac{\log (\eta(P_e))-\log \epsilon}{D_i+\epsilon}+\frac{1}{\mu_i D_i}+\EE[T]\\\nonumber
&= \frac{\eta(P_e)-\epsilon}{(I^i_L+\epsilon)}+ \frac{\log P_e + \log \frac{\eta(P_e)}{P_e}-\log \epsilon}{D_i+\epsilon}+\frac{1}{\mu_i D_i}+\EE[T]\\\label{eq: bound 2 on E[Z_T]}
&\leq \frac{\log P_e}{D_i+\epsilon} +\EE[T](1 + \delta_i(P_e, M_1M_2, \epsilon)),
\end{align}
where the function $\delta_i$ is defined as 
\begin{align*}
\delta_i(& P_e, M_1M_2, \epsilon)= \|  \frac{\eta(P_e)-\epsilon}{(I^i_L+\epsilon) \frac{\log M_1M_2}{R^{(3)}_N}}+ \frac{\log \frac{\eta(P_e)}{P_e}-\log \epsilon}{(D_i+\epsilon) \frac{\log M_1M_2}{R^{(3)}_N}}+\frac{1}{\mu_i D_i\frac{\log M_1M_2}{R^{(3)}_N}}\|
\end{align*}
Note that we use the equation $\EE[T]=\frac{\log M_1M_2}{R^{(3)}_N}$ in the definition of $\delta_i$. Observe that   $$\lim_{P_e\rightarrow 0 } \lim_{M_1M_2 \rightarrow \infty} \delta_i(P_e, M_1M_2, \epsilon)=0.$$ 
Note that $Z_0^i \leq \EE[Z_T^i], i=1,2,3$, where
$Z_0^i =  \frac{\log M_i-\epsilon}{I_L^i+\epsilon}.$
Therefore, 
\begin{align*}
\frac{\log M_i-\epsilon}{I_L^i+\epsilon} \leq \frac{\log P_e}{D_i+\epsilon}+\EE[T] ( 1 +  \delta_i(P_e, M_1M_2, \epsilon))
\end{align*}
Multiplying both sides by $\frac{D_i+\epsilon}{\EE[T]}$ and rearranging the terms give 

\begin{align*}
-\frac{\log P_e}{\EE[T]}& \leq  (D_i+\epsilon)\left(1-\frac{R^{(i)}_N}{I_L^i+\epsilon}\right)\\
&+\frac{\epsilon (D_i+\epsilon)}{(I_L^i+\epsilon)\EE[T]}+{(D_i+\epsilon)} \delta_i(P_e, M_1M_2, \epsilon),
\end{align*}

Define
\begin{align*}
\tilde{\delta}(P_e, M_1M_2, \epsilon)=\max_i \frac{(D_i+\epsilon) ~ \epsilon ~R^{(3)}_N}{(I_L^i+\epsilon') \log M_1M_2}+(D_i+\epsilon) \delta_i(P_e, M_1M_2, \epsilon).
\end{align*}

For any non-negative numbers $\lambda_i, i=1,2,3$ the following inequality holds:
\begin{align*}\numberthis \label{eq: bound on Exp as D_i}
-\frac{\log P_e}{\EE[T]}&  \leq  (D_i+\epsilon)\left(1- \frac{R^{(i)}_N}{I_L^i+\epsilon}\right)+\tilde{\delta},\\
&\leq  (D_i+\epsilon)\left(1-\frac{\lambda_i R^{(i)}_N}{\lambda_i I_L^i+\epsilon}\right)+\tilde{\delta},\\
&\leq  (D_i+\epsilon)\left(1-\frac{\lambda_i R^{(i)}_N}{\sum_j \lambda_j I_L^j+\epsilon'}\right)+\tilde{\delta},\\
&\leq  (D_i+\epsilon) \left(1-\frac{\lambda_i R^{(i)}_N}{\sup \sum_j \lambda_j I_L^j+\epsilon}\right)+\tilde{\delta},\\
&=  (D_i+\epsilon)\left(1-\frac{\lambda_i R^{(i)}_N}{C_{\lambda}+\epsilon}\right)+\tilde{\delta},
\end{align*}
Since the transmission rates are inside the capacity region, $\lambda_i R^{(i)}_N \leq C_{\lambda}$ and we obtain 
\begin{align*}
\frac{\log P_e}{\EE[T]}&  \leq  D_i\left(1-\frac{\lambda_i R^{(i)}_N}{C_{\lambda}+\epsilon}\right)+\epsilon+\tilde{\delta}(P_e, M_1M_2, \epsilon),\\
& \stackrel{(a)}{=}  D_i\left(1-\frac{\lambda_i R^{(i)}_N}{C_{\lambda}}\right)+ D_i \frac{\lambda_i R^{(i)}_N\epsilon}{C_{\lambda}(C_{\lambda}+\epsilon)} + \epsilon+\tilde{\delta}(P_e, M_1M_2, \epsilon),\\
& \stackrel{(b)}{\leq}  D_i\left(1-\frac{\lambda_i R^{(i)}_N}{C_{\lambda}}\right)+ D_{\max} \frac{\epsilon}{C_{\lambda}} + \epsilon+\tilde{\delta}(P_e, M_1M_2, \epsilon),\\
\end{align*}
where $D_{\max}=\max\{D_1, D_2, D_3\}$, (a) follows by adding and subtracting the term $D_i(\frac{\lambda_i R^{(i)}_N}{C_{\lambda}})$, and (b) follows as $\frac{\lambda_i R^{(i)}_N}{C_{\lambda}+\epsilon}\leq 1$. Define ${\delta}(P_e, M_1M_2, \epsilon)= \epsilon(1+  \frac{D_{\max}}{C_{\lambda}})+\tilde{\delta}(P_e, M_1M_2, \epsilon)$. The theorem follows by taking the minimum over $\lambda_i, i=1,2,3$ and the fact that the following condition is satisfied: $$\lim_{\epsilon \rightarrow 0}\lim_{P_e \rightarrow 0}\lim_{M_1M_2 \rightarrow \infty} \delta(P_e, M_1M_2, \epsilon)=0.$$ 
Note that in the above proof it is assumed that the capacity region is nonempty. This assumption implies that $C_{\lambda}>0$ for all $\underline{\lambda}\neq \underline{0}$ with non-negative components.   
\end{proof}

\section{Proof of Corollary \ref{cor: Exp upper bound}}\label{app: proof corrolary Exp upper bound}
From \eqref{eq: bound on Exp as D_i} in the proof of Theorem \ref{thm: Error Exp Upper bound}, we obtain: 
\begin{align*}
-\frac{\log P_e}{\EE[T]}&  \leq \min_{i \in \{1,2,3\}} (D_i+\epsilon)\left(1- \frac{R^{(i)}_N}{I_L^i+\epsilon}\right)+\tilde{\delta}\\
&\leq D_{\max}\min_{i \in \{1,2,3\}} \left(1- \frac{R^{(i)}_N}{I_L^i}\right)+\delta\\\numberthis \label{eq: cor bound i}
&= D_{\max} \min_{\substack{\alpha_1, \alpha_2, \alpha_3 \geq 0\\ \alpha_1+\alpha_2+\alpha_3=1}} \left(1- \sum_{i=1}^3 \alpha_i\frac{R^{(i)}_N}{I_L^i}\right)+{\delta},
\end{align*}
where $D_{\max}=\max\{D_1,D_2,D_3\}$, and $\delta=\tilde{\delta}+\epsilon \sup (1+\frac{D_{\max}}{I_L^i})$. For non-negative $\lambda_i, i=1,2,3$, set 
$\alpha_i = \frac{\lambda_i I_L^i}{\sum_j \lambda_j I_L^j}.$
Next, replace $\alpha_i, i=1,2,3$ in \eqref{eq: cor bound i} with the above term. Therefore, \eqref{eq: cor bound i} does not exceed the following
\begin{align*}
 &D_{\max} \min_{\substack{\lambda_1, \lambda_2, \lambda_3 \geq 0\\ \lambda_1+\lambda_2+\lambda_3=1}} \left(1-  \frac{\sum_{i}\lambda_i R^{(i)}_N}{\sum_j \lambda_j I_L^j}\right)+{\delta}.
\end{align*}
The proof is completed by noting that $\sum_j \lambda_j I_L^j \leq C_{\lambda} $.  

\end{document}